\documentclass[journal=nalefd,manuscript=article]{achemso}
\usepackage[version=3]{mhchem}
\usepackage{color}

\usepackage{epstopdf} 

\author{Biswanath Chakraborty}
\altaffiliation{Contributed equally to this work}
\affiliation{Department of Physics, Indian Institute of Science, Bangalore 560012, India}
\author{Satyendra Nath Gupta}
\altaffiliation{Contributed equally to this work}
\affiliation{Department of Physics, Indian Institute of Science, Bangalore 560012, India}
\author{Anjali Singh}
\affiliation{Theoretical Sciences Unit, Jawaharlal Nehru Centre for Advanced Scientific Research, Bangalore-560064, India}
\author{Manabendra Kuiri}
\affiliation{Department of Physics, Indian Institute of Science, Bangalore 560012, India}
\author{Chandan Kumar}
\affiliation{Department of Physics, Indian Institute of Science, Bangalore 560012, India}
\author{D. V. S. Muthu}
\affiliation{Department of Physics, Indian Institute of Science, Bangalore 560012, India}
\author{Anindya Das}
\affiliation{Department of Physics, Indian Institute of Science, Bangalore 560012, India}
\author{U. V. Waghmare}
\affiliation{Theoretical Sciences Unit, Jawaharlal Nehru Centre for Advanced Scientific Research, Bangalore-560064, India}
\author{{A. K. Sood}}
\affiliation{Department of Physics, Indian Institute of Science, Bangalore 560012, India}
\email{asood@physics.iisc.ernet.in}

\title{Electron-Hole Asymmetry in the Electron-phonon Coupling in Top-gated Phosphorene Transistor}


\begin{document}

\date{\today}

\begin{abstract}

Using in-situ Raman scattering from phosphorene channel in an electrochemically top-gated field effect transistor, we show that its phonons with A$_g$ symmetry depend much more strongly on concentration of electrons than that of holes, while the phonons with B$_g$ symmetry are insensitive to doping. With first-principles  theoretical analysis, we show that the observed electon-hole asymmetry arises from the radically different constitution of its conduction and valence bands involving $\pi$ and $\sigma$ bonding states respectively, whose symmetry permits coupling with only the phonons that preserve the lattice symmetry. Thus, Raman spectroscopy is a non-invasive tool for measuring electron concentration in phosphorene-based nanoelectronic devices.

\end{abstract}

\section{Introduction}

A few layer Black Phosphorus (BP) is a relatively new member of the family of 2D nanosystems with unique properties. Unlike gapless graphene, black phosphorus is a direct band gap material with gap ranging from 0.3 eV in bulk to 2 eV in monolayer,~\cite{liu2014phosphorene,PhysRevB.89.235319,castellanos2014isolation,xia2014rediscovering,wang2014highly,yang2015optical,liang2014electronic} covering a wide range of electromagnetic spectrum. This is again distinctive from transition metal dichalcogenides (TMD) which exhibit a direct band gap of $\sim$ 1 to 2 eV  only in their monolayer form. The narrow band gap of BP (in few layer form) bridges the gap between the zero gap graphene and large band gap TMDs, thus making BP a suitable candidate for near and mid infrared optics~\cite{buscema2014photovoltaic}. For possible applications in electronic devices, black phosphorus offers a good possibility with mobilitiy of $\sim$ 1000 cm$^2$/V-sec at room temperature  and On-Off ratio of $\sim$ 10$^5$ with excellent current saturation characteristics~\cite{li2014black}.

\begin{figure*}[ht!]
\begin{center}
\includegraphics [width=\textwidth]{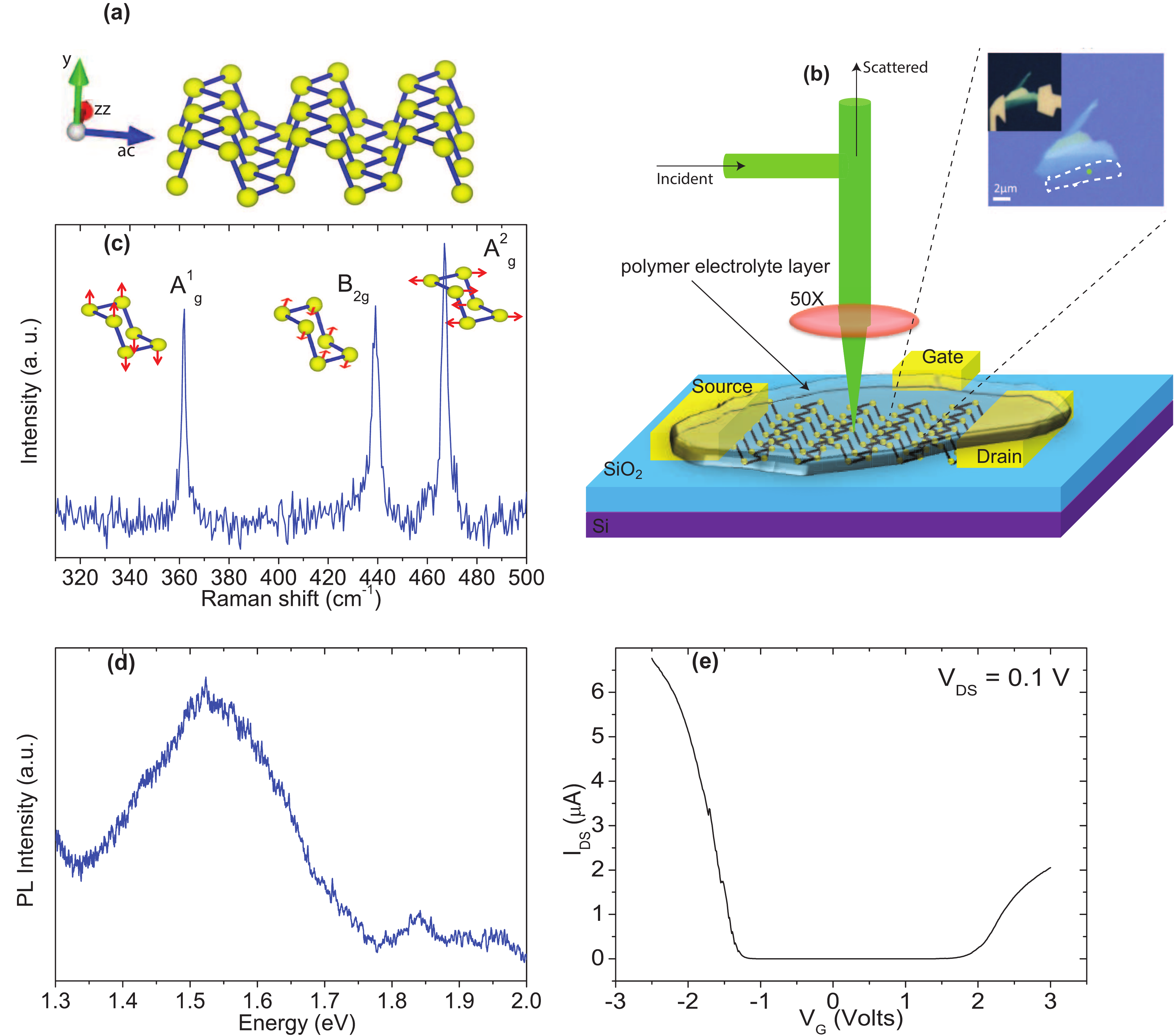}
\end{center}
\caption{\label{F1}(a)  Crystal structure of monolayer BP showing the armchair (ac) and zigzag (zz) directions. (b) A schematic of the experimental set-up, with inset showing optical image of the BP flake and the monolayer region is marked by a dashed line. (c)  Raman and (d) photoluminescence spectra recorded from the area dashed line. Laser wavelength used to record spectra is 532 nm. Inset of (c) shows the atomic displacements in normal modes. (e) Transport characteristics of the BP device.} 
\end{figure*}

Phosphorene (as monolayer BP is referred to) lattice structure is a 2-D puckered network (Figure \ref{F1}a) formed by phosphorus atoms with each atom covalently bonded to three neighbouring atoms. An unique feature of black phosphorus is its in-plane anisotropy. The puckered orthorhombic crystal structure of BP gives rise to asymmetric electronic and phonon dispersion relations: electronic bands are more dispersive along the armchair direction than along the zigzag direction, whereas the phonon dispersion is more dispersive along the zigzag direction~\cite{PhysRevB.89.235319,fei2014lattice,cai2015giant}. Along the armchair direction, carriers are more mobile, as the effective mass along the zigzag direction is about 10 times larger than that of along the armchair direction~\cite{akahama1983electrical,ling2015renaissance}. Similarly, the  Hall mobility along the armchair direction is 1.8 times the mobility along the zigzag direction ~\cite{xia2014rediscovering}. Single and few layers BP absorb light (ranging from infrared to part of the visible spectrum) polarized along the armchair axis, whereas they are transparent for polarization along the zigzag direction~\cite{PhysRevB.89.235319}.  Recent photoluminescence measurements on the monolayer show the existence of anisotropic excitons emitting light, polarized along the armchair direction, with a relatively higher exciton binding energy of around 0.7-0.8 eV and a quasi particle gap of 2.2 eV~\cite{wang2014highly,yang2015optical, PhysRevLett.115.026404,PhysRevB.90.075429}. 

For 2-D systems like graphene, electron-phonon coupling (EPC) plays a dominant role in the resistivity behaviour at high carrier densities~\cite{PhysRevLett.105.256805}. For almost a decade, since the emergence of graphene~\cite{novoselov2004electric, geim2007rise}, Raman spectroscopy has scored over other techniques as an indispensable analytical tool for estimation of the number of layers in 2D materials like graphene and TMDs~\cite{PhysRevLett.97.187401, gupta2006raman, graf2007spatially,lee2010anomalous,berkdemir2013identification},  to characterize EPC in 2-D graphene ~\cite{pisana2007breakdown, PhysRevLett.98.166802, das2008monitoring} and evaluating thermal conductivity of graphene layers~\cite{balandin2008superior,faugeras2010thermal,PhysRevB.83.081419} among various other properties. \textit{In-situ} Raman scattering from monolayer MoS$_2$ transistor have revealed the effect of doping on the EPC strength~\cite{PhysRevB.85.161403}.

 For BP too, Raman spectroscopy has shown its potential as a successful and noninvasive technique to determine the crystal orientation of black phosphorus flakes~\cite{xia2014rediscovering, wang2014highly,wu2015identifying, ribeiro2015unusual}. Recently Luo \textit{et. al.}~\cite{luo2015anisotropic} have reported anisotropic thermal conductivity using Raman spectroscopy: the thermal conductivity along the zigzag orientation exceeding that along the armchair direction by a factor of 2. The degradation of BP flakes due to environmental aspects have been probed by Raman spectroscopy~\cite{favron2015photooxidation}. With four atoms in the unit cell, out of nine optical modes, phosphorene has six Raman active modes, out of which three prominent modes with irreducible representations  A$^1$$_g$,  A$^2$$_g$ and  B$_{2g}$ are  observed in back scattering geometry. The eigen-vectors of these modes reveal that (Figure \ref{F1}c) A$^1$$_g$ mode involves out of plane displacements of atoms, while the atomic motions involved in A$^2$$_g$ and  B$_{2g}$ modes are along armchair and zigzag directions, respectively.

We identify and quantify the effects of doping on the Raman active phonons in phosphorene, revealing characteristic EPC. While A$^1$$_g$ and  A$^2$$_g$ phonons are affected by the doping, B$_{2g}$ mode is not. Further A$^1$$_g$,  A$^2$$_g$ modes are soften and their linewidths broaden with electron doping while remaining unaffected by holes. Experimental results are supported by first-principle density functional theory (DFT) calculations, which show an enhanced (reduced) EPC for A$^1$$_g$ and  A$^2$$_g$ modes with electron (hole) doping. We note that this is consistent with recent reports,~\cite{ge2015strain, shao2014electron} which suggest a possibility of phosphorene to be a BCS superconductor by virtue of its enhanced EPC at high concentration of electron doping and strain .

\section{Experimental Results and Discussions}

\begin{figure*}[ht!]
\begin{center}
\includegraphics [width=\textwidth]{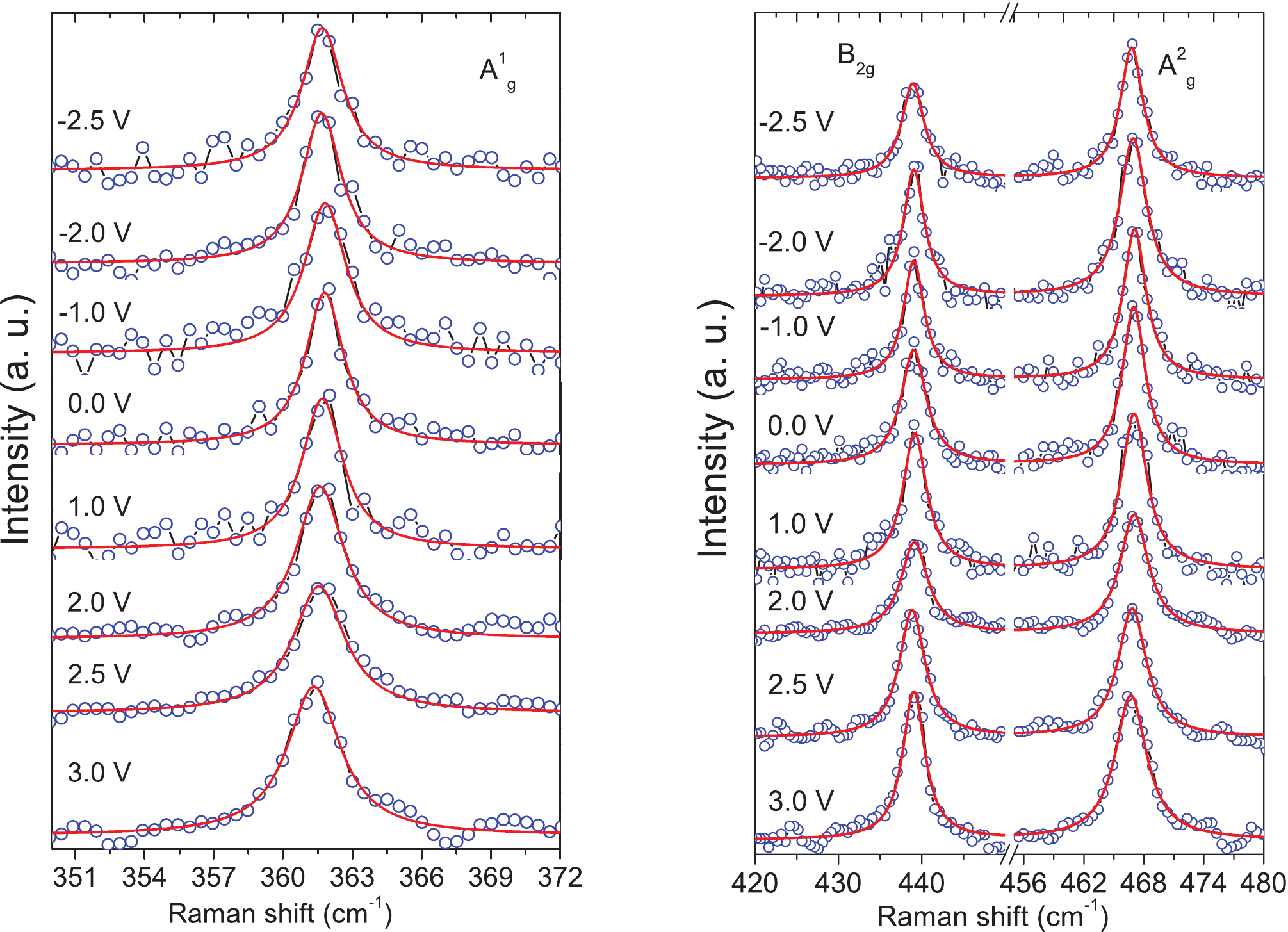}
\end{center}
\caption{\label{F2}  Raman spectra showing (a) A$^1$$_g$ and (b) B$_{2g}$ and A$^2$$_g$ modes for different top gate voltages. The top gate voltages are indicated in the figures. Circles are experimental data and lines are Lorentzian fit to the spectrum. The spectra were recorded from the spot marked by green dot in figure \ref{F1}b. }
\end{figure*}

Black phosphorus flakes were exfoliated from its single crystal (M/s Smart Elements), and transferred on to a 300 nm SiO$_2$  grown thermally on a heavily doped Si (M/s Nova Electronic Materials). Immediately after exfoliation,  the substrate was coated with a bilayer resist consisting of PMMA 450K/PMMA 950K to avoid degradation of black phosphorus  flakes~\cite{favron2015photooxidation}.  After careful observation under a microscope, flakes were selected and photoluminescence (PL) measurement were done to confirm the layer thickness. Figure \ref{F1}b shows the optical image of the selected flake. Raman spectrum showing the characteristic modes (namely A$^1$$_g$, B$_{2g}$ and A$^2$$_g$ ) and PL, recorded from the area marked by dashed line (Figure \ref{F1}b) are shown in  Figure \ref{F1}c and d respectively. The observed PL peak at $\sim$ 1.53 eV confirms the marked portion to be a single layer phosphorene\cite{xia2014rediscovering}. Source, drain and gate electrodes were fabricated by e-beam lithography followed by thermal evaporation of 5 nm/70 nm of Cr/Au and lift-off in acetone. A transparent solid polymer electrolyte gate consisting of lithium perchlorate (LiClO$_4$):polyethylene oxide (PEO) in the weight ratio of 1:8 is immediately drop casted to prevent the flake exposure. Due to the nanometer thick ($\sim$2 nm) Debye layer, this polymer electrolyte enables doping with higher carrier concentration (almost by an order of magnitude) as compared to 300 nm oxide gating~\cite{das2008monitoring}. Confocal Raman and PL measurements using 532 nm excitation laser are carried out at room temperature using Labram HR-800 coupled  with a Peltier cooled CCD and a spectrometer with 1800 lines/mm grating. Spectra are recorded using a long working distance 50X objective with numerical aperture 0.45. Spectra shown in Figure \ref{F2} were acquired for 10 seconds using a laser power of $\sim$ 300 $\mu$W.  Electrical measurements are done using Keithley 2400 source meters. Figure \ref{F1}b shows the schematic of the experimental set up and the inset shows the device. The device transfer characteristic is shown in Figure \ref{F1}e displaying an on-off ratio of $\sim$ 10$^5$ and a field effect mobility $\sim$ 100 cm$^2$/V-sec and $\sim$ 35 cm$^2$/V-sec for holes and electrons respectively. Similar asymmetric transfer characteristic curves were reported for the BP field effect transistors~\cite{li2014black,koenig2014electric,das2014tunable,cao2015quality}. In order to convert the applied gate voltage to carrier concentration $n$, we have used the standard MOSFET expression $n$ = C$_G$ (V$_G$ - V$_T$), where C$_G$, V$_G$ and V$_T$ are top gate capacitance, applied top gate voltage and threshold voltage respectively, and C$_G$ is taken to be 1.5$\mu$F/cm$^2$ ~\cite{chakraborty2009formation}. It may be noted that the threshold voltage could be different for an isolated monolayer device and can introduce some errors in estimating $n$. However, the qualitative picture and the  interpretation of the results would remain unaffected.

\begin{figure*}[ht!]
\begin{center}
\includegraphics [width=0.7\textwidth]{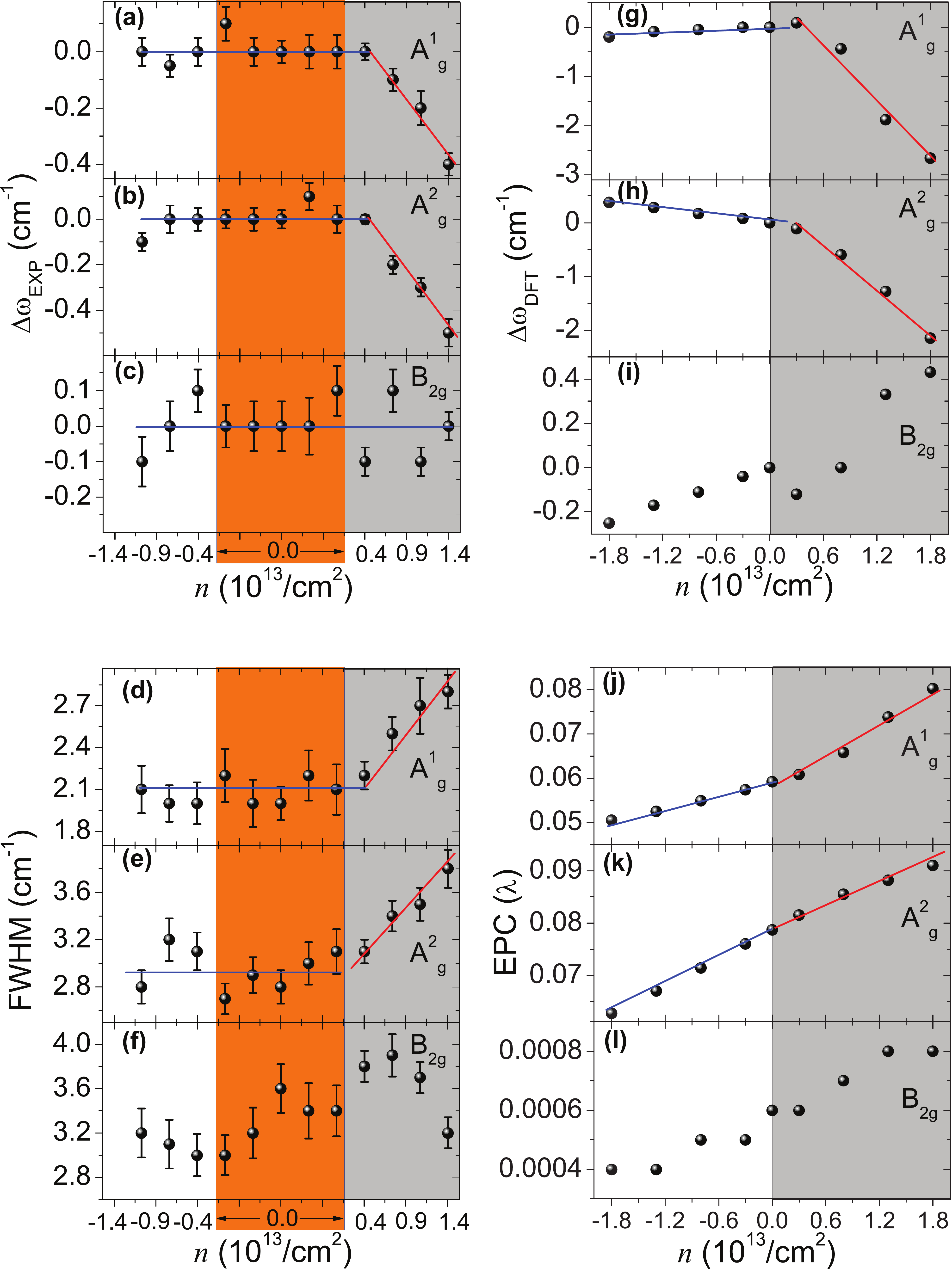}
\end{center}
\caption{\label{F3} (a) - (c) Changes in phonon frequencies $\Delta\omega$, and (d)- (f) FWHM of Raman peaks, as obtained from experiment, as a function of carrier concentration $n$. (g) - (i) $\Delta\omega$ obtained from DFT calculations and (j) - (l)  calculated values  of EPC ($\lambda$) as a function of  $n$. The lines are guide to eye. The off-state and electron doped regions are marked with orange and grey shades.}
\end{figure*}

 Figure \ref{F2} shows the evolution of the Raman modes of phosphorene as a function of gate voltage V$_G$. The peak positions and the full width at half maximum (FWHM) are obtained by fitting Lorentzian function (solid lines in Figure \ref{F2}) to the spectra. The changes in phonon energy $\Delta\omega$ = $\omega$($n$ $\neq$0) - $\omega$($n$=0)  are depicted in Figure \ref{F3}a, b and c. It can be seen that phonon softening is observed only for the A$^1$$_g$ and A$^2$$_g$ modes for electron doping(Figure \ref{F3} a, b). Hole doping has no effect on the phonon frequencies. As can be seen in Figure \ref{F3}c, the B$_{2g}$ mode is not affected by either electron or hole doping. Figure \ref{F3}d-f show FWHM of the three modes. The FWHM of A$^1$$_g$ and A$^2$$_g$ modes increase for the electron doping and do not change for the hole doping. The B$_{2g}$ linewidth does not show any significant doping dependence. Similar phonon and linewidth trends are observed for multilayer BP device (see Supplementary Information). We will now compare our results with the DFT calculations. 
 
  \begin{figure*}[ht!]
   \begin{center}
   \includegraphics [width=\textwidth]{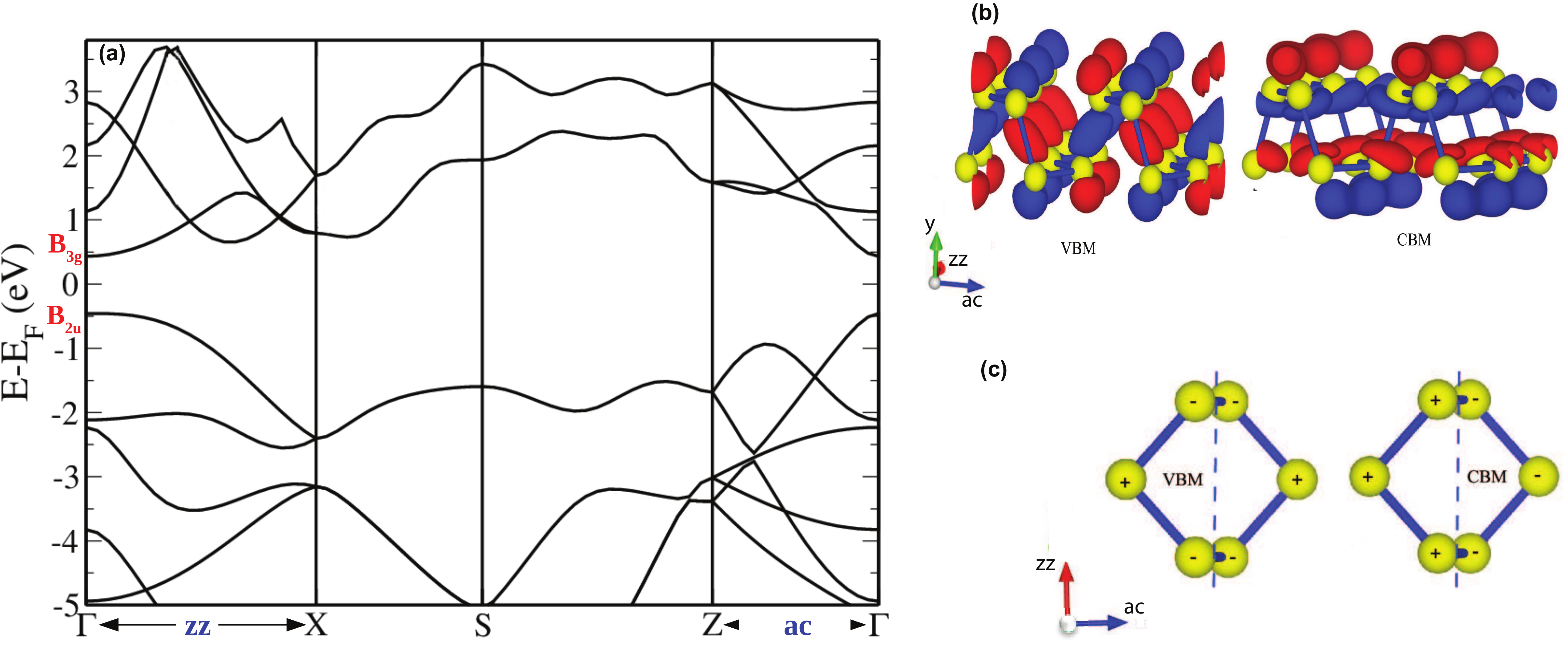}
   \end{center}
   \caption{\label{F4} (a) Electronic structure of phosphorene with symmetry labels B$_{2u}$ at VBM (valance band maximum) and B$_{3g}$ at CBM (conduction band minimum) at $\Gamma$-point, (b) isosurfaces of wavefunctions at VBM and CBM at the $\Gamma$-point, and (c) a schematic showing the symmetry of the wavefuctions at VBM and CBM at $\Gamma$-point where dashed line shows a mirror plane. Note that wavefuctions at VBM and CBM are even and odd respectively, under mirror reflection $\sigma$$_{ac}$.}
   \end{figure*}
   
    \begin{figure*}[ht!]
          \begin{center}
           \includegraphics [width=0.6\textwidth]{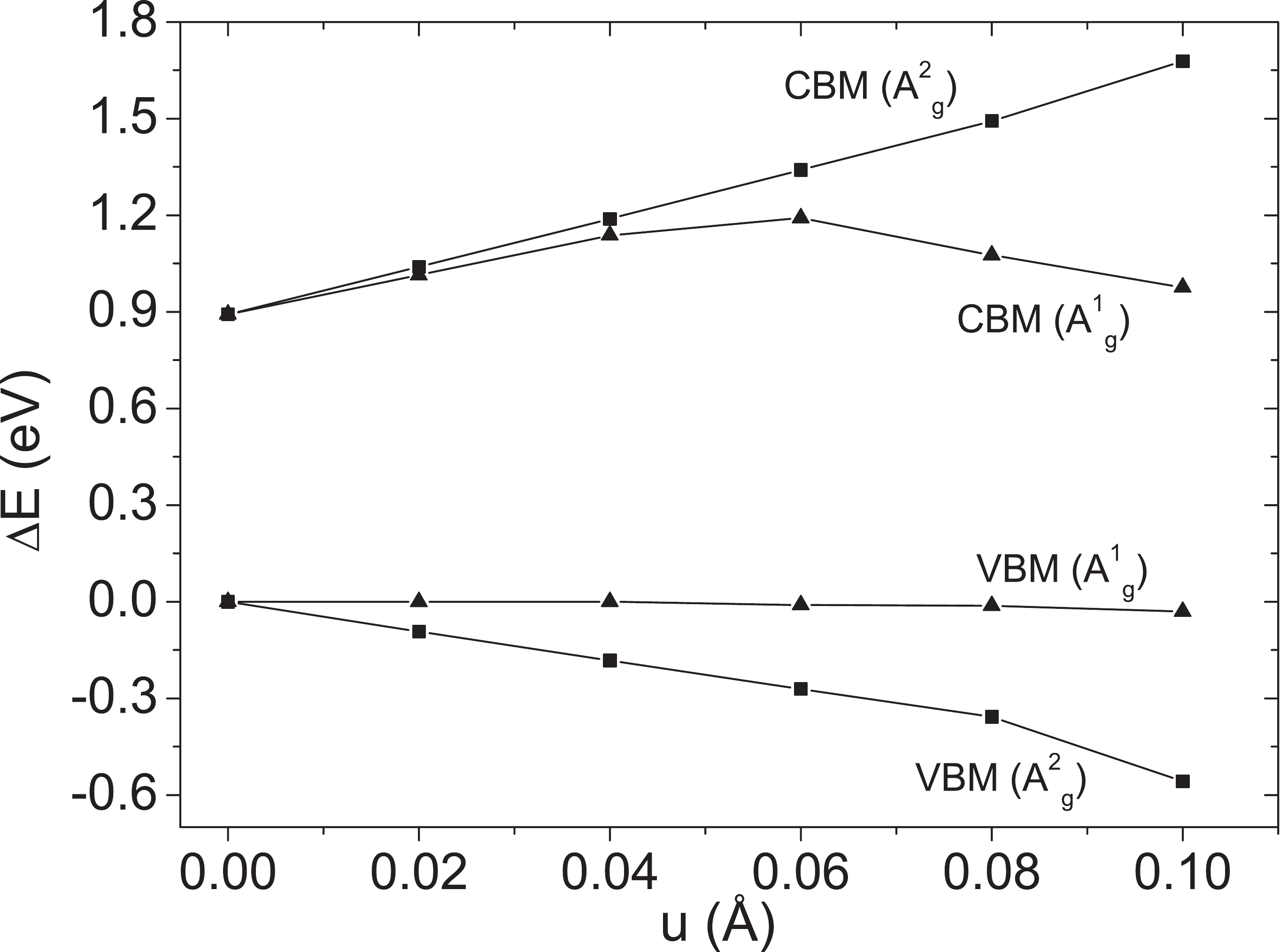}
          \end{center}
          \caption{\label{F5}Change in the energies of CBM and VBM as a function of structural distortion obtained by freezing in the atomic displacements of A$^1_g$ and A$^2_g$ modes.} 
          \end{figure*}

  \section{Theoretical Calculations}

 Our first-principles calculations are based on density functional theory as 
 implemented in Quantum ESPRESSO package,\cite{Giannozzi} and norm-conserving pseudopotentials \cite{HartwigsenC,GoedeckerS}
 to model the interaction between valence electrons and ionic cores.
 The exchange-correlation energy functional is treated within a Local Density Approximation (LDA) 
 with a parametrized form of Perdew and Zunger \cite{PerdewJP}.
 We use an energy cutoff of 60 Ry to truncate plane wave basis used 
 in representing Kohn-Sham wave functions, and an energy cutoff of 240 Ry 
 to represent charge density. Structures are relaxed till the Hellman-Feynman 
 force on each atom is less than 0.03 eV/\AA. We have used a periodic supercell 
 to simulate a phosphorene sheet, with a vacuum of 15 {\AA} (along y-direction) to separate its adjacent periodic images.
 Integrations over the Brillouin Zone (BZ) are sampled on 12x1x9 and 48x1x36 uniform meshes of k-points 
 in the calculation of total energy and electron-phonon coupling respectively.
 For a few concentrations of hole doping (1.3 x 10$^{13}$ cm$^{-2}$ and 1.8 x 10$^{13}$ cm$^{-2}$), we used 
 72x1x54 uniform mesh of k-points in the determination of electron-phonon coupling to ensure convergence.

 \par Electronic structure of phosphorene (single layer BP) determined from our first-principles calculations exhibits a direct band gap
 of 0.89 eV (Figure \ref{F4}a) at the $\Gamma$ point, in good agreement with earlier theoretical results \cite{liu2014phosphorene}. 
 We note that the electronic structure is remarkably different along zigzag ($\Gamma$X) and armchair ($\Gamma$Z) directions: linear 
 dispersion along the armchair (ac) and parabolic dispersion along zigzag (zz) directions. Electronic states at conduction band 
 minimum (CBM) and valance band maximum (VBM) at $\Gamma$ point  consist of p$_y$ orbitals of phosphorous (Figure \ref{F4}b). 
 We note that symmetries of the wavefunctions at CBM and VBM are distinct (see Figure \ref{F4}c), which lead to contrasting effects of electron and hole doping on the electron-phonon coupling (to be explained later).

 \par We have simulated carrier (electron and hole) doping in phosphorene by adding a small fraction of electrons/holes (according to doping concentration) to the unit cell. From the changes in frequencies of the A$^{1}_{g}$ and A$^{2}_{g}$ modes (Figure \ref{F3}g-i), it is clear that A$^{1}_{g}$ and A$^{2}_{g}$ modes  soften significantly with electron doping (A$^{1}_{g}$ $\sim$ 3 cm$^{-1}$ and A$^{2}_{g}$ $\sim$ 2 cm$^{-1}$ at ~ $1.8\times10^{13}$ cm$^{-2}$) as compared to insignificant change with hole doping. In contrast, B$_{2g}$ is much less affected by doping. These results are in qualitative agreement with our experimental observations (see Figure \ref{F3}a-c). 
 To understand these trends, we systematically determined electron-phonon coupling with carrier doping. 
 The electron-phonon coupling (EPC) of a phonon mode $\nu$ at wavevector q (frequeccy $\omega$) is ~\cite{attaccalite2010doped, PhysRevB.85.161403}
 
 \begin{eqnarray}\nonumber 
 \label{lambda}
 \lambda_{\mbox{\scriptsize\boldmath $q$}\nu} &=&
 \frac{2}{\hbar\omega_{\mbox{\scriptsize\boldmath $q$}\nu}N(\epsilon_F)}\sum_{\mbox{\scriptsize\boldmath $k$}}\sum_{ij}|g_{\mbox{\scriptsize\boldmath $k+ q,k$}}^{\mbox{\scriptsize\boldmath $q$}\nu ,ij}|^2\times\delta(\epsilon_{\mbox{\scriptsize\boldmath $k+ q$},i}-\epsilon_F) \\
 &{\times}& 
 \delta(\epsilon_{\mbox{\scriptsize\boldmath $k$},j}-\epsilon_F),
 \end{eqnarray}
  where N($\epsilon_{F}$) is the density of states at Fermi energy. 
 The electron-phonon coupling matrix element is 
 
\begin{equation}
\label{g}
g_{\mbox{\scriptsize\boldmath $k + q,k$}}^{\mbox{\scriptsize\boldmath $q$}\nu ,ij}=\left(\frac {\hbar}{2M\omega_{\mbox{\scriptsize\boldmath $q$}\nu}}\right)^\frac{1}{2}\langle\psi_{\mbox{\scriptsize\boldmath $k + q$},i}|\triangle V_{\mbox{\scriptsize\boldmath $q$}\nu}|\psi_{\mbox{\scriptsize\boldmath $k$},j}\rangle,
\end{equation}
 where $\psi_{k,j}$ is electronic wavefunction of j$^{th}$ band  at wavevector k, M is the effective mass associated with the phonon, and $\triangle V_{\mbox{\scriptsize\boldmath $q$}\nu}$ =
 $\frac{\partial V}{\partial u^{\nu}(q)}$ is the change in the self-consistent potential associated with atomic 
 displacements of phonon q$\nu$.

  From calculated $\lambda$ for all the Raman modes as a function of carrier concentration (Figure \ref{F3}j-l), it is seen that (a) A$^1_g$ and A$^2_g$ modes couple more strongly with the carriers than the B$_{2g}$ mode, and (b) all the modes 
 couple more strongly with electrons than with holes. These results are indeed consistent with our experiments. Strong coupling of the A$^{1}_{g}$ 
 mode with electrons can be understood from symmetry arguments. A$_g$ modes (see Figure \ref{F1}c) have 
 the full symmetry of the lattice, (\textit{i.e.} distortion along these modes does not break the symmetry of the lattice) and  
hence all the electronic states (i = j) have a non-zero matrix element (Eq.2). 
 Using the same argument, matrix element (i = j in Eq.2) vanishes for the B$_{2g}$ mode which breaks the symmetry of the 
 lattice (B$_{2g}$ mode is orthogonal to the A$_g$ modes). We do find an asymmetry in the EPC with
 hole and electron doping: EPC increases with increasing concentration of electron doping while it decreases 
 with increasing hole concentration. This asymmetry follows from the symmetry of the electronic wavefunctions at the VBM and 
 CBM at $\Gamma$ point. At the VBM, wavefunction is even under mirror reflection, while it is odd at the CBM. As a result, the magnitude as well as the sign of the coupling matrix (Eq.2) can be different for electron and hole doping. 
 It is evident in Figure \ref{F3}g-h and \ref{F3}j-k that A$^1_g$ mode is affected more strongly by electron doping than the hole doping. In contrast,  A$^2_g$ mode is affected by both electron as well as by hole doping, though the effect is stronger for electrons. To understand this, we obtained the energies of VBM and CBM at $\Gamma$ point as a function of structural distortions obtained by freezing the eigenvectors of these modes.  It is evident (Figure \ref{F5} ) that the CBM varies more strongly than the VBM with A$^1_g$ mode, whereas freezing of A$^2_g$ mode (Figure\ref{F5}) affects both VBM and CBM significantly, though its effect on the CBM is slightly stronger than that on the VBM. Therefore, the difference in the dependence of A$^1_g$ and A$^2_g$ modes on electron and hole doping originates  from the matrix element $\langle\psi_{k,i}|\Delta V_{q\nu}|\psi_{k,j}\rangle$.  The non-monotonous evolution of CBM as a function of distortion for A$^1_g$ mode is further explained in the supplementary material.
 
 We note that low levels of doping results in changes in the population of only the states that have energies close to the VBM and CBM at $\Gamma$ point. Thus, the states lining the band-gap along $\Gamma$-X direction are more relevant to our experiments than the ones along $\Gamma$-Z (energies of the states at Z-point are too far (Fig. 4a) from the gap to be affected by low levels of doping in phosphorene transistor). In Fig. S5a, we have  shown iso-surfaces of wavefunctions of the valence and conduction bands at X-point and a diagram of their projections on p-orbitals of P.  We note that the bands at X point are doubly degenerate, and are made from p$_x$ and p$_y$ orbitals of P, in contrast to the VBM and CBM at $\Gamma$ point made from p$_y$ orbitals. From the Fig. S5b, it is clear that the relative phases of p orbitals of the P sublattices are the same for wave functions at X and $\Gamma$ (Fig. 4c) points. Hence, our symmetry analysis of the coupling of Raman active phonons to the states along $\Gamma$ to X is similar. Since the orbitals involved in the VBM and CBM bands at X point are different from those at $\Gamma$ point, their relative phase factors are not simple. In fact, they involve rotations in the 3-dimensional space of bands made of two p-orbitals. Thus, their relative phases can be meaningfully described with a Hermitian matrix that is responsible for the complex Raman tensor relevant to the polarization dependent Raman scattering\cite{ribeiro2015unusual}. To connect more closely with the results of reference 32, we determined optical dielectric constants of structures obtained by freezing A$^1_g$ and A$^2_g$ modes. $\Delta\varepsilon_{xx}$   and  $\Delta\varepsilon_{zz}$  are 0.05 and 1.3 for A$^1_g$ and 0.2 and 1.5 for A$^2_g$ modes, respectively. This confirms  a remarkable anisotropy in the Raman tensor, quite distinct for A$^1_g$ and A$^2_g$ modes, in qualitative agreement with the fitted parameters obtained from polarization dependent Raman spectra of Phosphorene \cite{ribeiro2015unusual}.

 \section{Conclusions}
 
 Our work demonstrates a phosphorene-based field effect transistor (P-FET) with a high on-off ratio, in which the carrier concentration in phosphorene channel can be effectively controlled by the top-gate electrode. We have shown how \textit{in-situ} Raman spectroscopy can be a powerful tool in assessing the nature and concentration of carriers in the channel of P-FET. While the anisotropy in its properties is expected from its structural symmetry, we uncover a remarkable electron-hole asymmetry in the coupling between its charge carriers and phonons: phonons with A$_g$ symmetry couple much more strongly with electrons than with holes. Further, only the phonons preserving the symmetry of the lattice couple strongly with electrons. Our first-principles calculations reveal that the electron-hole asymmetry arises from rather different characters of conduction and valence bands involving  $\pi$ and $\sigma$ bonding states respectively.

The angular dependence of the intensities of A$^1_g$, A$^2_g$  and B2g phonons in black phosphorus (BP) has been recently well studied\cite{ribeiro2015unusual}. It is shown that for reproducing experimental results for intensities of Ag modes, the Raman tensor has to be complex whereas for B$_{2g}$, real values of Raman tensor elements can explain the angular dependence. The complex values of the Raman tensor can arise both from electron-radiation as well as electron-phonon matrix elements. An interesting aspect brought out  is that the imaginary part of the Raman tensor elements is different for the two totally symmetric modes A$^1_g$ and A$^2_g$ , pointing out that electron-phonon interaction is a key factor responsible for the complex values of the tensor. In our work, we have shown that electron-phonon coupling (EPC) associated with A$^1_g$ and A$^2_g$ modes vary predominantly with electron doping. Our experiments did not focus on the angular dependence of the Raman intensities. The work of Ref. 32 by Ribeiro et.al. points out an interesting future possibility to study the angular dependence of Raman intensities of A$_g$ modes as a function of carrier concentration and see how the relative phase factor (i.e. the imaginary part of the Raman tensor elements) for the A$_g$ mode varies. This will further quantify the contribution of electron-phonon matrix elements to the real and imaginary parts of the Raman tensor elements. Our work is fundamental to understanding the transport in phosphorene and key to measurement of carriers in phosphorene-based nanoelectronic devices.

\begin{acknowledgement}
BC, AKS, and UVW thank the financial support under the Nano Mission project of the Department of Science and Technology, Government of India. AD thanks IISc start up grant for the financial support.
\end{acknowledgement}

\providecommand*\mcitethebibliography{\thebibliography}
\csname @ifundefined\endcsname{endmcitethebibliography}
  {\let\endmcitethebibliography\endthebibliography}{}


\begin{mcitethebibliography}{46}
\providecommand*\natexlab[1]{#1}
\providecommand*\mciteSetBstSublistMode[1]{}
\providecommand*\mciteSetBstMaxWidthForm[2]{}
\providecommand*\mciteBstWouldAddEndPuncttrue
  {\def\EndOfBibitem{\unskip.}}
\providecommand*\mciteBstWouldAddEndPunctfalse
  {\let\EndOfBibitem\relax}
\providecommand*\mciteSetBstMidEndSepPunct[3]{}
\providecommand*\mciteSetBstSublistLabelBeginEnd[3]{}
\providecommand*\EndOfBibitem{}
\mciteSetBstSublistMode{f}
\mciteSetBstMaxWidthForm{subitem}{(\alph{mcitesubitemcount})}
\mciteSetBstSublistLabelBeginEnd
  {\mcitemaxwidthsubitemform\space}
  {\relax}
  {\relax}

\bibitem[Liu et~al.(2014)Liu, Neal, Zhu, Luo, Xu, Tom{\'a}nek, and
  Ye]{liu2014phosphorene}
Liu,~H.; Neal,~A.~T.; Zhu,~Z.; Luo,~Z.; Xu,~X.; Tom{\'a}nek,~D.; Ye,~P.~D.
  \emph{ACS Nano} \textbf{2014}, \emph{8}, 4033--4041\relax
\mciteBstWouldAddEndPuncttrue
\mciteSetBstMidEndSepPunct{\mcitedefaultmidpunct}
{\mcitedefaultendpunct}{\mcitedefaultseppunct}\relax
\EndOfBibitem
\bibitem[Tran et~al.(2014)Tran, Soklaski, Liang, and Yang]{PhysRevB.89.235319}
Tran,~V.; Soklaski,~R.; Liang,~Y.; Yang,~L. \emph{Phys. Rev. B} \textbf{2014},
  \emph{89}, 235319\relax
\mciteBstWouldAddEndPuncttrue
\mciteSetBstMidEndSepPunct{\mcitedefaultmidpunct}
{\mcitedefaultendpunct}{\mcitedefaultseppunct}\relax
\EndOfBibitem
\bibitem[Castellanos-Gomez et~al.(2014)Castellanos-Gomez, Vicarelli, Prada,
  Island, Narasimha-Acharya, Blanter, Groenendijk, Buscema, Steele, Alvarez,
  Zandbergen, Palacios, and van~der Zant]{castellanos2014isolation}
Castellanos-Gomez,~A.; Vicarelli,~L.; Prada,~E.; Island,~J.~O.;
  Narasimha-Acharya,~K.; Blanter,~S.~I.; Groenendijk,~D.~J.; Buscema,~M.;
  Steele,~G.~A.; Alvarez,~J.; Zandbergen,~H.~W.; Palacios,~J.; van~der Zant,~H.
  S.~J. \emph{2D Materials} \textbf{2014}, \emph{1}, 025001\relax
\mciteBstWouldAddEndPuncttrue
\mciteSetBstMidEndSepPunct{\mcitedefaultmidpunct}
{\mcitedefaultendpunct}{\mcitedefaultseppunct}\relax
\EndOfBibitem
\bibitem[Xia et~al.(2014)Xia, Wang, and Jia]{xia2014rediscovering}
Xia,~F.; Wang,~H.; Jia,~Y. \emph{Nature Communications} \textbf{2014},
  \emph{5}, 4458\relax
\mciteBstWouldAddEndPuncttrue
\mciteSetBstMidEndSepPunct{\mcitedefaultmidpunct}
{\mcitedefaultendpunct}{\mcitedefaultseppunct}\relax
\EndOfBibitem
\bibitem[Wang et~al.(2015)Wang, Jones, Seyler, Tran, Jia, Zhao, Wang, Yang, Xu,
  and Xia]{wang2014highly}
Wang,~X.; Jones,~A.~M.; Seyler,~K.~L.; Tran,~V.; Jia,~Y.; Zhao,~H.; Wang,~H.;
  Yang,~L.; Xu,~X.; Xia,~F. \emph{Nature Nanotechnology} \textbf{2015},
  \emph{10}, 517--521\relax
\mciteBstWouldAddEndPuncttrue
\mciteSetBstMidEndSepPunct{\mcitedefaultmidpunct}
{\mcitedefaultendpunct}{\mcitedefaultseppunct}\relax
\EndOfBibitem
\bibitem[Yang et~al.(2015)Yang, Xu, Pei, Myint, Wang, Wang, Zhang, Yu, and
  Lu]{yang2015optical}
Yang,~J.; Xu,~R.; Pei,~J.; Myint,~Y.~W.; Wang,~F.; Wang,~Z.; Zhang,~S.; Yu,~Z.;
  Lu,~Y. \emph{Light:Science and Application} \textbf{2015}, \emph{4}\relax
\mciteBstWouldAddEndPuncttrue
\mciteSetBstMidEndSepPunct{\mcitedefaultmidpunct}
{\mcitedefaultendpunct}{\mcitedefaultseppunct}\relax
\EndOfBibitem
\bibitem[Liang et~al.(2014)Liang, Wang, Lin, Sumpter, Meunier, and
  Pan]{liang2014electronic}
Liang,~L.; Wang,~J.; Lin,~W.; Sumpter,~B.~G.; Meunier,~V.; Pan,~M. \emph{Nano
  Letters} \textbf{2014}, \emph{14}, 6400--6406\relax
\mciteBstWouldAddEndPuncttrue
\mciteSetBstMidEndSepPunct{\mcitedefaultmidpunct}
{\mcitedefaultendpunct}{\mcitedefaultseppunct}\relax
\EndOfBibitem
\bibitem[Buscema et~al.(2014)Buscema, Groenendijk, Steele, van~der Zant, and
  Castellanos-Gomez]{buscema2014photovoltaic}
Buscema,~M.; Groenendijk,~D.~J.; Steele,~G.~A.; van~der Zant,~H.~S.;
  Castellanos-Gomez,~A. \emph{Nature Communications} \textbf{2014}, \emph{5},
  4651\relax
\mciteBstWouldAddEndPuncttrue
\mciteSetBstMidEndSepPunct{\mcitedefaultmidpunct}
{\mcitedefaultendpunct}{\mcitedefaultseppunct}\relax
\EndOfBibitem
\bibitem[Li et~al.(2014)Li, Yu, Ye, Ge, Ou, Wu, Feng, Chen, and
  Zhang]{li2014black}
Li,~L.; Yu,~Y.; Ye,~G.~J.; Ge,~Q.; Ou,~X.; Wu,~H.; Feng,~D.; Chen,~X.~H.;
  Zhang,~Y. \emph{Nature Nanotechnology} \textbf{2014}, \emph{9},
  372--377\relax
\mciteBstWouldAddEndPuncttrue
\mciteSetBstMidEndSepPunct{\mcitedefaultmidpunct}
{\mcitedefaultendpunct}{\mcitedefaultseppunct}\relax
\EndOfBibitem
\bibitem[Fei and Yang(2014)Fei, and Yang]{fei2014lattice}
Fei,~R.; Yang,~L. \emph{Applied Physics Letters} \textbf{2014}, \emph{105},
  083120\relax
\mciteBstWouldAddEndPuncttrue
\mciteSetBstMidEndSepPunct{\mcitedefaultmidpunct}
{\mcitedefaultendpunct}{\mcitedefaultseppunct}\relax
\EndOfBibitem
\bibitem[Cai et~al.(2015)Cai, Ke, Zhang, Feng, Shenoy, and Zhang]{cai2015giant}
Cai,~Y.; Ke,~Q.; Zhang,~G.; Feng,~Y.~P.; Shenoy,~V.~B.; Zhang,~Y.-W.
  \emph{Advanced Functional Materials} \textbf{2015}, \emph{25},
  2230--2236\relax
\mciteBstWouldAddEndPuncttrue
\mciteSetBstMidEndSepPunct{\mcitedefaultmidpunct}
{\mcitedefaultendpunct}{\mcitedefaultseppunct}\relax
\EndOfBibitem
\bibitem[Akahama et~al.(1983)Akahama, Endo, and Narita]{akahama1983electrical}
Akahama,~Y.; Endo,~S.; Narita,~S.-i. \emph{Journal of the Physical Society of
  Japan} \textbf{1983}, \emph{52}, 2148--2155\relax
\mciteBstWouldAddEndPuncttrue
\mciteSetBstMidEndSepPunct{\mcitedefaultmidpunct}
{\mcitedefaultendpunct}{\mcitedefaultseppunct}\relax
\EndOfBibitem
\bibitem[Ling et~al.(2015)Ling, Wang, Huang, Xia, and
  Dresselhaus]{ling2015renaissance}
Ling,~X.; Wang,~H.; Huang,~S.; Xia,~F.; Dresselhaus,~M.~S. \emph{Proceedings of
  the National Academy of Sciences} \textbf{2015}, \emph{112},
  4523?4530\relax
\mciteBstWouldAddEndPuncttrue
\mciteSetBstMidEndSepPunct{\mcitedefaultmidpunct}
{\mcitedefaultendpunct}{\mcitedefaultseppunct}\relax
\EndOfBibitem
\bibitem[Schuster et~al.(2015)Schuster, Trinckauf, Habenicht, Knupfer, and
  B\"uchner]{PhysRevLett.115.026404}
Schuster,~R.; Trinckauf,~J.; Habenicht,~C.; Knupfer,~M.; B\"uchner,~B.
  \emph{Phys. Rev. Lett.} \textbf{2015}, \emph{115}, 026404\relax
\mciteBstWouldAddEndPuncttrue
\mciteSetBstMidEndSepPunct{\mcitedefaultmidpunct}
{\mcitedefaultendpunct}{\mcitedefaultseppunct}\relax
\EndOfBibitem
\bibitem[Rodin et~al.(2014)Rodin, Carvalho, and
  Castro~Neto]{PhysRevB.90.075429}
Rodin,~A.~S.; Carvalho,~A.; Castro~Neto,~A.~H. \emph{Phys. Rev. B}
  \textbf{2014}, \emph{90}, 075429\relax
\mciteBstWouldAddEndPuncttrue
\mciteSetBstMidEndSepPunct{\mcitedefaultmidpunct}
{\mcitedefaultendpunct}{\mcitedefaultseppunct}\relax
\EndOfBibitem
\bibitem[Efetov and Kim(2010)Efetov, and Kim]{PhysRevLett.105.256805}
Efetov,~D.~K.; Kim,~P. \emph{Phys. Rev. Lett.} \textbf{2010}, \emph{105},
  256805\relax
\mciteBstWouldAddEndPuncttrue
\mciteSetBstMidEndSepPunct{\mcitedefaultmidpunct}
{\mcitedefaultendpunct}{\mcitedefaultseppunct}\relax
\EndOfBibitem
\bibitem[Novoselov et~al.(2004)Novoselov, Geim, Morozov, Jiang, Zhang, Dubonos,
  Grigorieva, and Firsov]{novoselov2004electric}
Novoselov,~K.~S.; Geim,~A.~K.; Morozov,~S.; Jiang,~D.; Zhang,~Y.; Dubonos,~S.;
  Grigorieva,~I.; Firsov,~A. \emph{science} \textbf{2004}, \emph{306},
  666--669\relax
\mciteBstWouldAddEndPuncttrue
\mciteSetBstMidEndSepPunct{\mcitedefaultmidpunct}
{\mcitedefaultendpunct}{\mcitedefaultseppunct}\relax
\EndOfBibitem
\bibitem[Geim and Novoselov(2007)Geim, and Novoselov]{geim2007rise}
Geim,~A.~K.; Novoselov,~K.~S. \emph{Nature Materials} \textbf{2007}, \emph{6},
  183--191\relax
\mciteBstWouldAddEndPuncttrue
\mciteSetBstMidEndSepPunct{\mcitedefaultmidpunct}
{\mcitedefaultendpunct}{\mcitedefaultseppunct}\relax
\EndOfBibitem
\bibitem[Ferrari et~al.(2006)Ferrari, Meyer, Scardaci, Casiraghi, Lazzeri,
  Mauri, Piscanec, Jiang, Novoselov, Roth, and Geim]{PhysRevLett.97.187401}
Ferrari,~A.~C.; Meyer,~J.~C.; Scardaci,~V.; Casiraghi,~C.; Lazzeri,~M.;
  Mauri,~F.; Piscanec,~S.; Jiang,~D.; Novoselov,~K.~S.; Roth,~S.; Geim,~A.~K.
  \emph{Phys. Rev. Lett.} \textbf{2006}, \emph{97}, 187401\relax
\mciteBstWouldAddEndPuncttrue
\mciteSetBstMidEndSepPunct{\mcitedefaultmidpunct}
{\mcitedefaultendpunct}{\mcitedefaultseppunct}\relax
\EndOfBibitem
\bibitem[Gupta et~al.(2006)Gupta, Chen, Joshi, Tadigadapa, and
  Eklund]{gupta2006raman}
Gupta,~A.; Chen,~G.; Joshi,~P.; Tadigadapa,~S.; Eklund,~P. \emph{Nano Letters}
  \textbf{2006}, \emph{6}, 2667--2673\relax
\mciteBstWouldAddEndPuncttrue
\mciteSetBstMidEndSepPunct{\mcitedefaultmidpunct}
{\mcitedefaultendpunct}{\mcitedefaultseppunct}\relax
\EndOfBibitem
\bibitem[Graf et~al.(2007)Graf, Molitor, Ensslin, Stampfer, Jungen, Hierold,
  and Wirtz]{graf2007spatially}
Graf,~D.; Molitor,~F.; Ensslin,~K.; Stampfer,~C.; Jungen,~A.; Hierold,~C.;
  Wirtz,~L. \emph{Nano Letters} \textbf{2007}, \emph{7}, 238--242\relax
\mciteBstWouldAddEndPuncttrue
\mciteSetBstMidEndSepPunct{\mcitedefaultmidpunct}
{\mcitedefaultendpunct}{\mcitedefaultseppunct}\relax
\EndOfBibitem
\bibitem[Lee et~al.(2010)Lee, Yan, Brus, Heinz, Hone, and
  Ryu]{lee2010anomalous}
Lee,~C.; Yan,~H.; Brus,~L.~E.; Heinz,~T.~F.; Hone,~J.; Ryu,~S. \emph{ACS nano}
  \textbf{2010}, \emph{4}, 2695--2700\relax
\mciteBstWouldAddEndPuncttrue
\mciteSetBstMidEndSepPunct{\mcitedefaultmidpunct}
{\mcitedefaultendpunct}{\mcitedefaultseppunct}\relax
\EndOfBibitem
\bibitem[Berkdemir et~al.(2013)Berkdemir, Guti{\'e}rrez, Botello-M{\'e}ndez,
  Perea-L{\'o}pez, El{\'\i}as, Chia, Wang, Crespi, L{\'o}pez-Ur{\'\i}as,
  Charlier, Terrones, and Terrones]{berkdemir2013identification}
Berkdemir,~A.; Guti{\'e}rrez,~H.~R.; Botello-M{\'e}ndez,~A.~R.;
  Perea-L{\'o}pez,~N.; El{\'\i}as,~A.~L.; Chia,~C.-I.; Wang,~B.; Crespi,~V.~H.;
  L{\'o}pez-Ur{\'\i}as,~F.; Charlier,~J.-C.; Terrones,~H.; Terrones,~M.
  \emph{Scientific reports} \textbf{2013}, \emph{3}\relax
\mciteBstWouldAddEndPuncttrue
\mciteSetBstMidEndSepPunct{\mcitedefaultmidpunct}
{\mcitedefaultendpunct}{\mcitedefaultseppunct}\relax
\EndOfBibitem
\bibitem[Pisana et~al.(2007)Pisana, Lazzeri, Casiraghi, Novoselov, Geim,
  Ferrari, and Mauri]{pisana2007breakdown}
Pisana,~S.; Lazzeri,~M.; Casiraghi,~C.; Novoselov,~K.~S.; Geim,~A.~K.;
  Ferrari,~A.~C.; Mauri,~F. \emph{Nature Materials} \textbf{2007}, \emph{6},
  198--201\relax
\mciteBstWouldAddEndPuncttrue
\mciteSetBstMidEndSepPunct{\mcitedefaultmidpunct}
{\mcitedefaultendpunct}{\mcitedefaultseppunct}\relax
\EndOfBibitem
\bibitem[Yan et~al.(2007)Yan, Zhang, Kim, and Pinczuk]{PhysRevLett.98.166802}
Yan,~J.; Zhang,~Y.; Kim,~P.; Pinczuk,~A. \emph{Phys. Rev. Lett.} \textbf{2007},
  \emph{98}, 166802\relax
\mciteBstWouldAddEndPuncttrue
\mciteSetBstMidEndSepPunct{\mcitedefaultmidpunct}
{\mcitedefaultendpunct}{\mcitedefaultseppunct}\relax
\EndOfBibitem
\bibitem[Das et~al.(2008)Das, Pisana, Chakraborty, Piscanec, Saha, Waghmare,
  Novoselov, Krishnamurthy, Geim, Ferrari, and Sood]{das2008monitoring}
Das,~A.; Pisana,~S.; Chakraborty,~B.; Piscanec,~S.; Saha,~S.; Waghmare,~U.;
  Novoselov,~K.; Krishnamurthy,~H.; Geim,~A.; Ferrari,~A.; Sood,~A.
  \emph{Nature Nanotechnology} \textbf{2008}, \emph{3}, 210--215\relax
\mciteBstWouldAddEndPuncttrue
\mciteSetBstMidEndSepPunct{\mcitedefaultmidpunct}
{\mcitedefaultendpunct}{\mcitedefaultseppunct}\relax
\EndOfBibitem
\bibitem[Balandin et~al.(2008)Balandin, Ghosh, Bao, Calizo, Teweldebrhan, Miao,
  and Lau]{balandin2008superior}
Balandin,~A.~A.; Ghosh,~S.; Bao,~W.; Calizo,~I.; Teweldebrhan,~D.; Miao,~F.;
  Lau,~C.~N. \emph{Nano Letters} \textbf{2008}, \emph{8}, 902--907\relax
\mciteBstWouldAddEndPuncttrue
\mciteSetBstMidEndSepPunct{\mcitedefaultmidpunct}
{\mcitedefaultendpunct}{\mcitedefaultseppunct}\relax
\EndOfBibitem
\bibitem[Faugeras et~al.(2010)Faugeras, Faugeras, Orlita, Potemski, Nair, and
  Geim]{faugeras2010thermal}
Faugeras,~C.; Faugeras,~B.; Orlita,~M.; Potemski,~M.; Nair,~R.~R.; Geim,~A.
  \emph{ACS Nano} \textbf{2010}, \emph{4}, 1889--1892\relax
\mciteBstWouldAddEndPuncttrue
\mciteSetBstMidEndSepPunct{\mcitedefaultmidpunct}
{\mcitedefaultendpunct}{\mcitedefaultseppunct}\relax
\EndOfBibitem
\bibitem[Lee et~al.(2011)Lee, Yoon, Kim, Lee, and Cheong]{PhysRevB.83.081419}
Lee,~J.-U.; Yoon,~D.; Kim,~H.; Lee,~S.~W.; Cheong,~H. \emph{Phys. Rev. B}
  \textbf{2011}, \emph{83}, 081419\relax
\mciteBstWouldAddEndPuncttrue
\mciteSetBstMidEndSepPunct{\mcitedefaultmidpunct}
{\mcitedefaultendpunct}{\mcitedefaultseppunct}\relax
\EndOfBibitem
\bibitem[Chakraborty et~al.(2012)Chakraborty, Bera, Muthu, Bhowmick, Waghmare,
  and Sood]{PhysRevB.85.161403}
Chakraborty,~B.; Bera,~A.; Muthu,~D. V.~S.; Bhowmick,~S.; Waghmare,~U.~V.;
  Sood,~A.~K. \emph{Phys. Rev. B} \textbf{2012}, \emph{85}, 161403\relax
\mciteBstWouldAddEndPuncttrue
\mciteSetBstMidEndSepPunct{\mcitedefaultmidpunct}
{\mcitedefaultendpunct}{\mcitedefaultseppunct}\relax
\EndOfBibitem
\bibitem[Wu et~al.()Wu, Mao, Xie, Xu, and Zhang]{wu2015identifying}
Wu,~J.; Mao,~N.; Xie,~L.; Xu,~H.; Zhang,~J. \emph{Angewandte Chemie} \emph{54},
  2366--2369\relax
\mciteBstWouldAddEndPuncttrue
\mciteSetBstMidEndSepPunct{\mcitedefaultmidpunct}
{\mcitedefaultendpunct}{\mcitedefaultseppunct}\relax
\EndOfBibitem
\bibitem[Ribeiro et~al.(2015)Ribeiro, Pimenta, de~Matos, Moreira, Rodin,
  Zapata, de~Souza, and Castro~Neto]{ribeiro2015unusual}
Ribeiro,~H.~B.; Pimenta,~M.~A.; de~Matos,~C.~J.; Moreira,~R.~L.; Rodin,~A.~S.;
  Zapata,~J.~D.; de~Souza,~E.~A.; Castro~Neto,~A.~H. \emph{ACS Nano}
  \textbf{2015}, \emph{9}, 4270?4276\relax
\mciteBstWouldAddEndPuncttrue
\mciteSetBstMidEndSepPunct{\mcitedefaultmidpunct}
{\mcitedefaultendpunct}{\mcitedefaultseppunct}\relax
\EndOfBibitem
\bibitem[Luo et~al.(2015)Luo, Maassen, Deng, Du, Lundstrom, Ye, and
  Xu]{luo2015anisotropic}
Luo,~Z.; Maassen,~J.; Deng,~Y.; Du,~Y.; Lundstrom,~M.~S.; Ye,~P.~D.; Xu,~X.
  \emph{Nature Communications} \textbf{2015}, \relax
\mciteBstWouldAddEndPunctfalse
\mciteSetBstMidEndSepPunct{\mcitedefaultmidpunct}
{}{\mcitedefaultseppunct}\relax
\EndOfBibitem
\bibitem[Favron et~al.(2015)Favron, Gaufr{\`e}s, Fossard, Phaneuf-L?Heureux,
  Tang, L{\'e}vesque, Loiseau, Leonelli, Francoeur, and
  Martel]{favron2015photooxidation}
Favron,~A.; Gaufr{\`e}s,~E.; Fossard,~F.; Phaneuf-L?Heureux,~A.-L.;
  Tang,~N.~Y.; L{\'e}vesque,~P.~L.; Loiseau,~A.; Leonelli,~R.; Francoeur,~S.;
  Martel,~R. \emph{Nature Materials} \textbf{2015}, \emph{14}, 826--832\relax
\mciteBstWouldAddEndPuncttrue
\mciteSetBstMidEndSepPunct{\mcitedefaultmidpunct}
{\mcitedefaultendpunct}{\mcitedefaultseppunct}\relax
\EndOfBibitem
\bibitem[Ge et~al.(2015)Ge, Wan, Yang, and Yao]{ge2015strain}
Ge,~Y.; Wan,~W.; Yang,~F.; Yao,~Y. \emph{New Journal of Physics} \textbf{2015},
  \emph{17}, 035008\relax
\mciteBstWouldAddEndPuncttrue
\mciteSetBstMidEndSepPunct{\mcitedefaultmidpunct}
{\mcitedefaultendpunct}{\mcitedefaultseppunct}\relax
\EndOfBibitem
\bibitem[Shao et~al.(2014)Shao, Lu, Lv, and Sun]{shao2014electron}
Shao,~D.; Lu,~W.; Lv,~H.; Sun,~Y. \emph{EPL (Europhysics Letters)}
  \textbf{2014}, \emph{108}, 67004\relax
\mciteBstWouldAddEndPuncttrue
\mciteSetBstMidEndSepPunct{\mcitedefaultmidpunct}
{\mcitedefaultendpunct}{\mcitedefaultseppunct}\relax
\EndOfBibitem
\bibitem[Koenig et~al.(2014)Koenig, Doganov, Schmidt, Neto, and
  Oezyilmaz]{koenig2014electric}
Koenig,~S.~P.; Doganov,~R.~A.; Schmidt,~H.; Neto,~A.~C.; Oezyilmaz,~B.
  \emph{Applied Physics Letters} \textbf{2014}, \emph{104}, 103106\relax
\mciteBstWouldAddEndPuncttrue
\mciteSetBstMidEndSepPunct{\mcitedefaultmidpunct}
{\mcitedefaultendpunct}{\mcitedefaultseppunct}\relax
\EndOfBibitem
\bibitem[Das et~al.(2014)Das, Zhang, Demarteau, Hoffmann, Dubey, and
  Roelofs]{das2014tunable}
Das,~S.; Zhang,~W.; Demarteau,~M.; Hoffmann,~A.; Dubey,~M.; Roelofs,~A.
  \emph{Nano Letters} \textbf{2014}, \emph{14}, 5733--5739\relax
\mciteBstWouldAddEndPuncttrue
\mciteSetBstMidEndSepPunct{\mcitedefaultmidpunct}
{\mcitedefaultendpunct}{\mcitedefaultseppunct}\relax
\EndOfBibitem
\bibitem[Cao et~al.(2015)Cao, Mishchenko, Yu, Khestanova, Rooney, Prestat,
  Kretinin, Blake, Shalom, Balakrishnan, Grigorieva, Novoselov, Piot, Potemski,
  Watanabe, taniguchi, Haigh, Geim, and Gorbachev]{cao2015quality}
Cao,~Y. et~al.  \emph{Nano Letters} \textbf{2015}, \emph{15}, 4914--4921\relax
\mciteBstWouldAddEndPuncttrue
\mciteSetBstMidEndSepPunct{\mcitedefaultmidpunct}
{\mcitedefaultendpunct}{\mcitedefaultseppunct}\relax
\EndOfBibitem
\bibitem[Chakraborty et~al.(2009)Chakraborty, Das, and
  Sood]{chakraborty2009formation}
Chakraborty,~B.; Das,~A.; Sood,~A. \emph{Nanotechnology} \textbf{2009},
  \emph{20}, 365203\relax
\mciteBstWouldAddEndPuncttrue
\mciteSetBstMidEndSepPunct{\mcitedefaultmidpunct}
{\mcitedefaultendpunct}{\mcitedefaultseppunct}\relax
\EndOfBibitem
\bibitem[Giannozzi et~al.(2009)Giannozzi, Baroni, Bonini, Calandra, Car,
  Cavazzoni, Ceresoli, Chiarotti, Cococcioni, Dabo, Corso, de~Gironcoli,
  Fabris, Fratesi, Gebauer, Gerstmann, Gougoussis, Kokalj, Lazzeri,
  Martin-Samos, Marzari, Mauri, Mazzarello, Paolini, Pasquarello, Paulatto,
  Sbraccia, Scandolo, Sclauzero, PSeitsonen, Smogunov, Umari, and
  Wentzcovitch]{Giannozzi}
Giannozzi,~P. et~al.  \emph{J. Phys.: Condens. Matter} \textbf{2009},
  \emph{21}, 395502\relax
\mciteBstWouldAddEndPuncttrue
\mciteSetBstMidEndSepPunct{\mcitedefaultmidpunct}
{\mcitedefaultendpunct}{\mcitedefaultseppunct}\relax
\EndOfBibitem
\bibitem[Hartwigsen et~al.(1998)Hartwigsen, Goedecker, and Hutter]{HartwigsenC}
Hartwigsen,~C.; Goedecker,~S.; Hutter,~J. \emph{Phys. Rev. B} \textbf{1998},
  \emph{58}, 3641--3662\relax
\mciteBstWouldAddEndPuncttrue
\mciteSetBstMidEndSepPunct{\mcitedefaultmidpunct}
{\mcitedefaultendpunct}{\mcitedefaultseppunct}\relax
\EndOfBibitem
\bibitem[Goedecker et~al.(1996)Goedecker, Teter, and Hutter]{GoedeckerS}
Goedecker,~S.; Teter,~M.; Hutter,~J. \emph{Phys. Rev. B} \textbf{1996},
  \emph{54}, 1703--1710\relax
\mciteBstWouldAddEndPuncttrue
\mciteSetBstMidEndSepPunct{\mcitedefaultmidpunct}
{\mcitedefaultendpunct}{\mcitedefaultseppunct}\relax
\EndOfBibitem
\bibitem[Perdew and Zunger(1981)Perdew, and Zunger]{PerdewJP}
Perdew,~J.~P.; Zunger,~A. \emph{Phys. Rev. B} \textbf{1981}, \emph{23},
  5048--5079\relax
\mciteBstWouldAddEndPuncttrue
\mciteSetBstMidEndSepPunct{\mcitedefaultmidpunct}
{\mcitedefaultendpunct}{\mcitedefaultseppunct}\relax
\EndOfBibitem
\bibitem[Attaccalite et~al.(2010)Attaccalite, Wirtz, Lazzeri, Mauri, and
  Rubio]{attaccalite2010doped}
Attaccalite,~C.; Wirtz,~L.; Lazzeri,~M.; Mauri,~F.; Rubio,~A. \emph{Nano
  Letters} \textbf{2010}, \emph{10}, 1172--1176\relax
\mciteBstWouldAddEndPuncttrue
\mciteSetBstMidEndSepPunct{\mcitedefaultmidpunct}
{\mcitedefaultendpunct}{\mcitedefaultseppunct}\relax
\EndOfBibitem
\end{mcitethebibliography}

\clearpage
\newpage
\begin{center}

\end{center}
\title{\bf{SUPPLEMENTARY INFORMATION}\\\bf{Electron-Hole Asymmetry in the Electron-phonon Coupling in Top-gated Phosphorene Transistor}}

\renewcommand\thefigure{SF\arabic{figure}}

\section{Experimental Results for the Multilayer BP}
\begin{figure}[tbh]
\includegraphics[width=\textwidth]{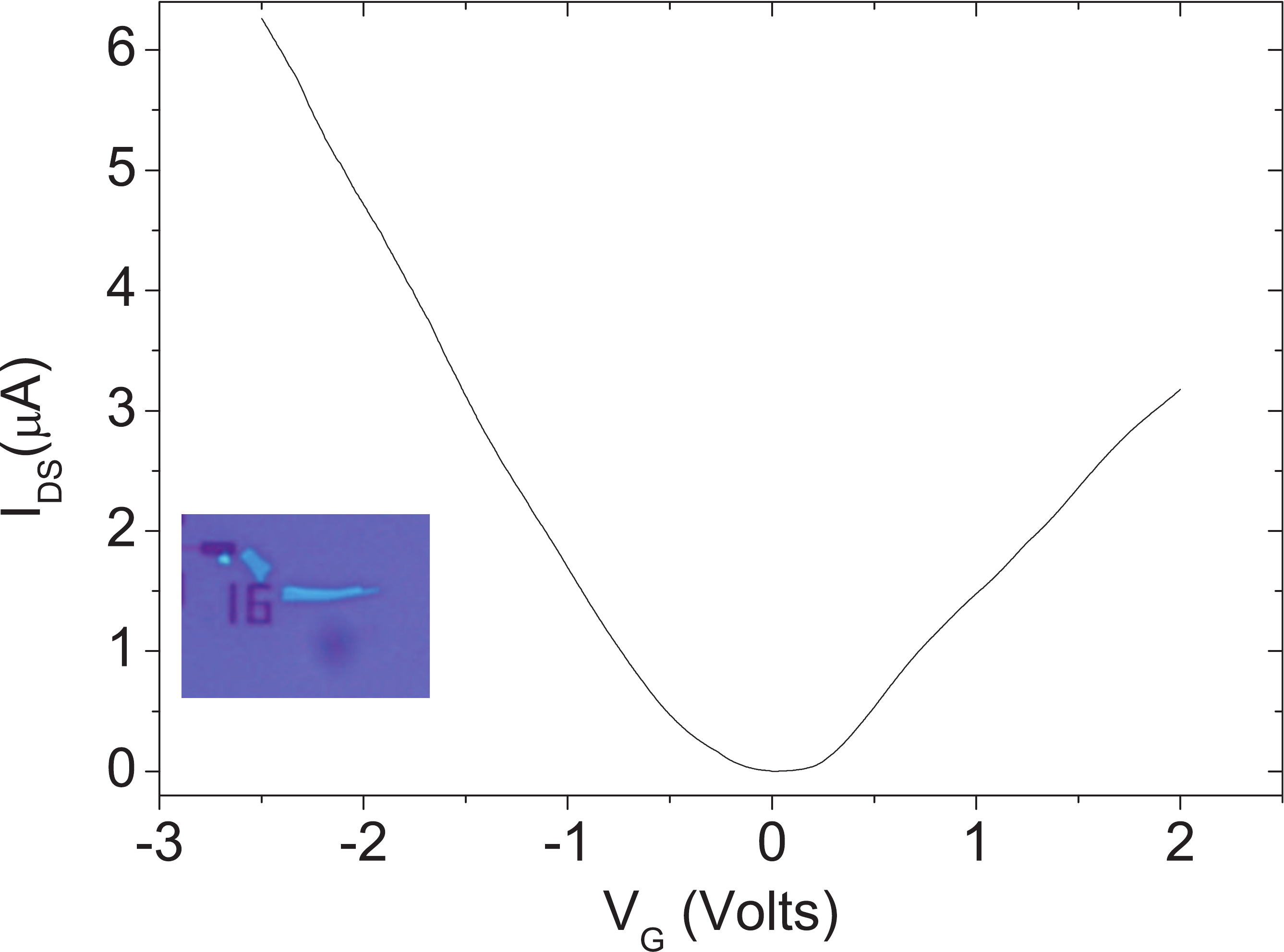}
\small{\caption{\label{S1} Multilayer BP transfer characteristic. Drain-source volatge was 0.05 V. Inset shows the multilayer flake.}}
\end{figure}

\begin{figure}[tbh]
\includegraphics[width=\textwidth]{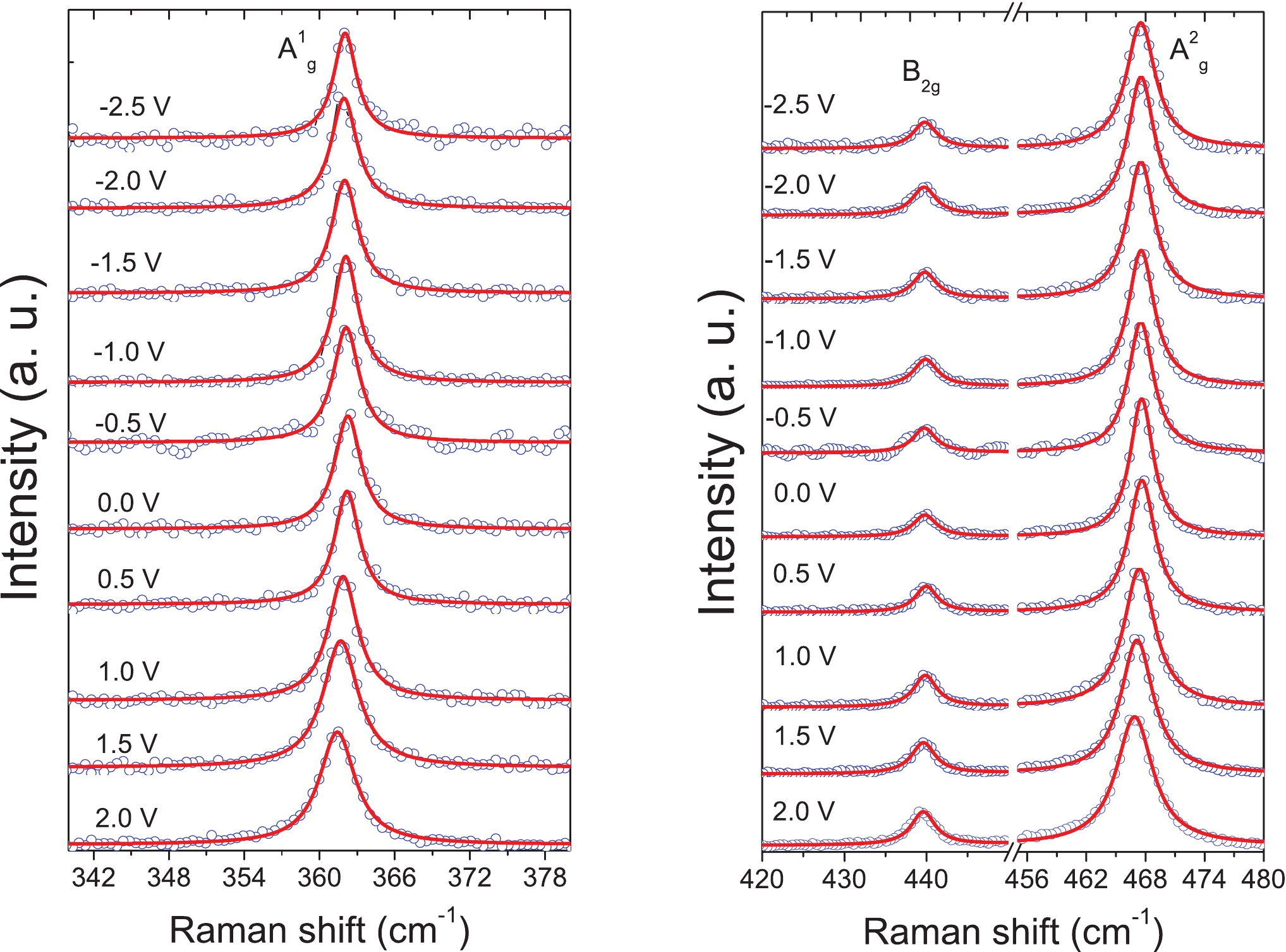}
\small{\caption{\label{S2} Raman spectra from multilayer BP showing (a) A$^1$$_g$ and (b) B$_{2g}$ and A$^1$$_g$ modes for different top gate voltages. The top gate voltages are indicated in the figures. Circles are experimental data and lines are Lorentzian fit to the spectrum.}}
\end{figure}

\begin{figure}[tbh]
\includegraphics[width=\textwidth]{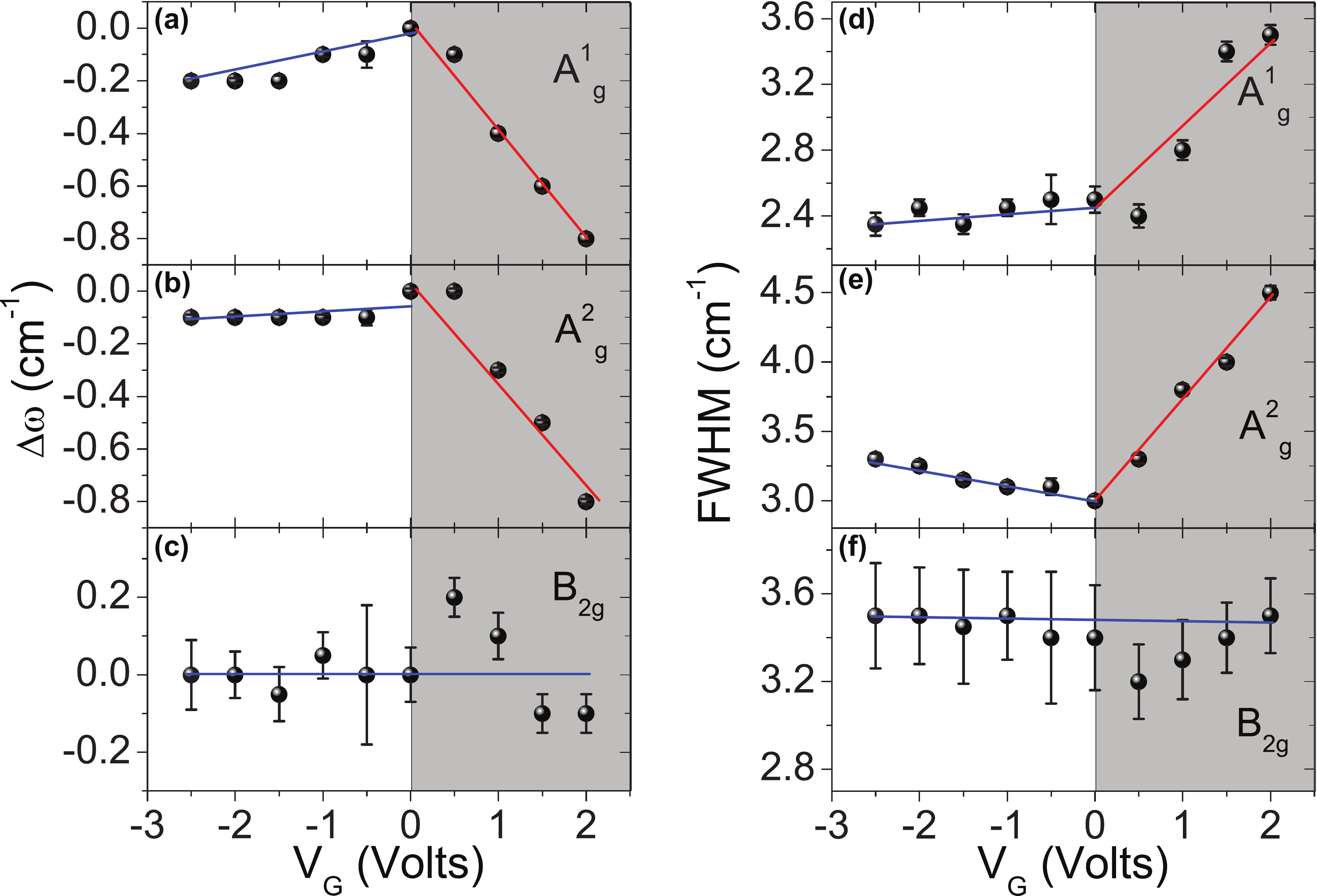}
\small{\caption{\label{S3} (a)- (c) Changes in Phonon frequency $\Delta\omega$, as a function of carrier concentration $n$. (d) - (f) FWHM as a function of $n$. Electron doped region is shaded. The lines are guide to eye.}}
\end{figure}

We observe similar trends for A$_g$ and B$_{2g}$ modes  for a top-gated multilayer BP ($\sim$ 9 layers) device. The transfer characteristics are shown in figure \ref{S1}. The field effect mobilities were estimated to be 40 cm$^2$/V-sec and 25 cm$^2$/V-sec for holes and electrons, respectively. Raman spectra showing the evolution of different modes at different top gate voltages are shown in figure \ref{S2} a and b. The frequency and linewidth pattern are similar to what we have demonstrated for monolayer. For doping with electrons, the A$^1$$_g$ and A$^2$$_g$ modes soften (figure \ref{S3} a and b) while remaining unchanged for hole doping. The FWHM for these modes increase for electron side only (figure \ref{S3} d and e). The B$_{2g}$ mode is not affected by gating as evident from frequency and linewidth trends in figure \ref{S3} c and f, respectively.

\section{Electronic Structure of Phosphorene}
\begin{figure}[tbh]
\includegraphics[width=0.8\textwidth]{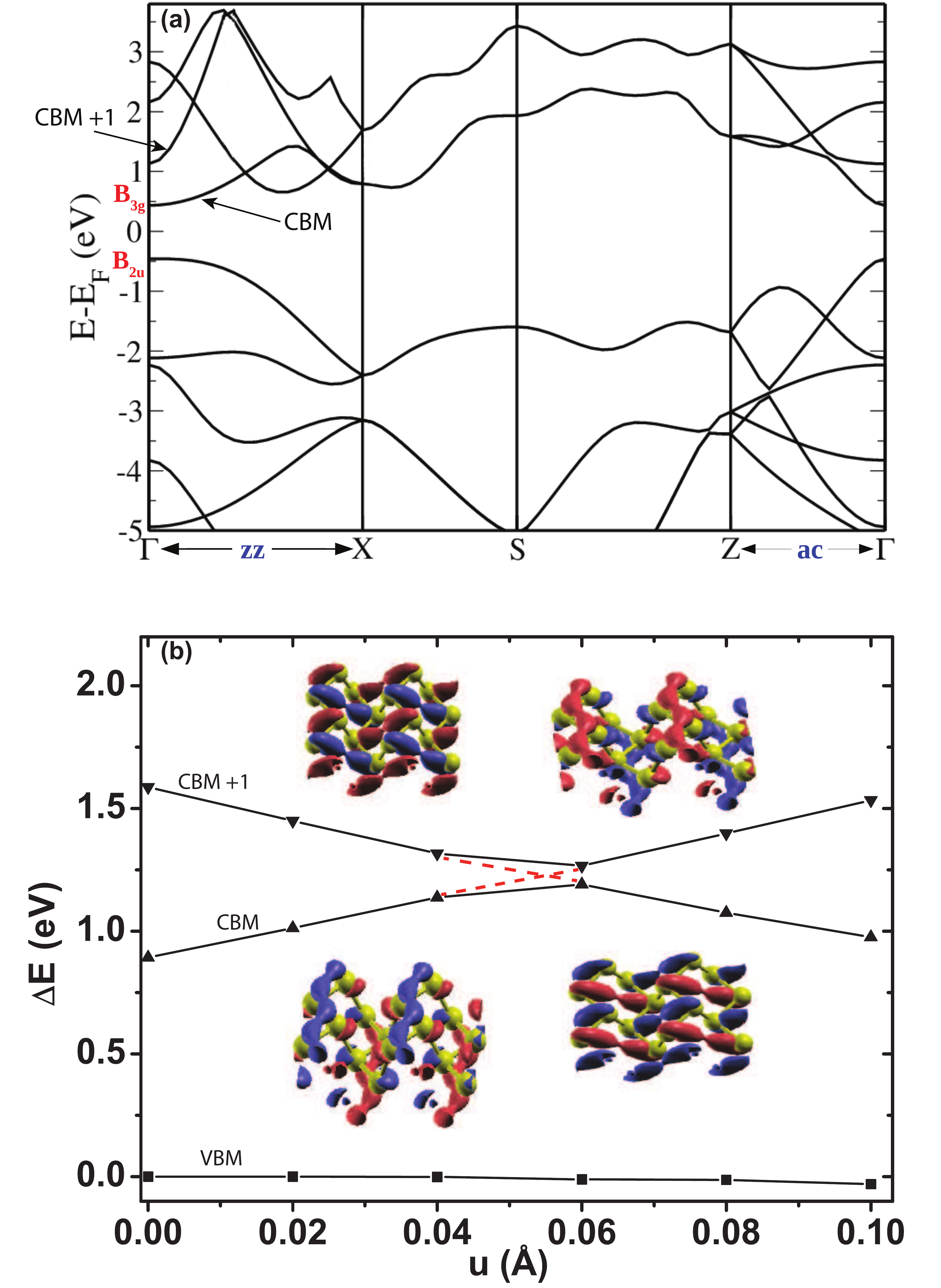}
\small{\caption{\label{S4} (a) Electronic structure of phosphorene with CBM and CBM +1 marked.  (b) Isosurfaces of wavefunctions of CBM and CBM+1 with structural distortion of A$^1_g$ mode for 0.04 {\AA} and 0.06 {\AA}, that reveal the band inversion between CBM and CBM+1 bands. If both the p$_y$ orbitals are seen in the same reference frame across the red dashed line,  we observe that p$_y$ orbitals get flipped by 180$^0$ i.e. $\psi$ = -$\psi$ confirms band inversion of CBM and CBM+1.}}
\end{figure}

\begin{figure}[tbh]
\includegraphics[width=0.8\textwidth]{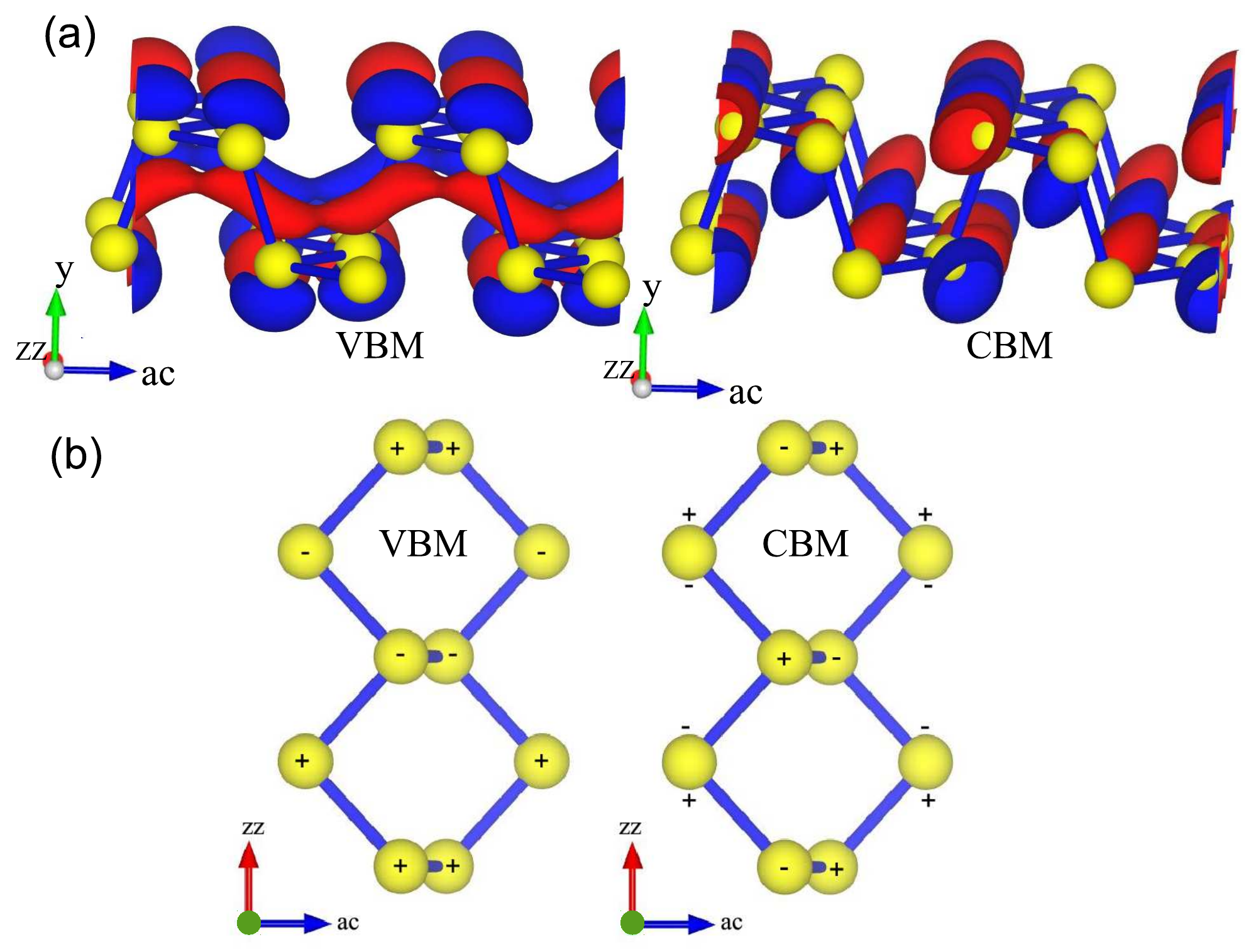}
\small{\caption{\label{S5} (a) Isosurfaces of wavefunctions at VBM and CBM at the X-point and (b) a schematic diagram of projection on p$_{x,y}$-orbitals of P showing symmetry of the wavefunction at VBM and CBM at X-point. Note that states at the VBM and CBM at X-point are doubly degenerate and isosurfaces of the degenerate states (VBM and VBM-1 or CBM and CBM+1) are similar.}}  
\end{figure}

Figure \ref{S4} a shows electronic structure of phosphorene calculated using DFT where the conduction bands CBM and CBM+1 are marked.
Figure \ref{S4} b shows the isosurfaces of wavefunctions of CBM and CBM+1 with structural distortion of A$^1_g$ mode for 0.04 {\AA} and 0.06 {\AA}, that reveal the band inversion between CBM and CBM+1 bands. The p$_y$ orbitals get flip by 180$^0$ s i.e. $\psi$ = -$\psi$ confirming band inversion of CBM and CBM+1. This can be the reason for the non-monotonous dependence of CBM energy on the structural distortion associated with the A$^1_g$ mode shown in Figure 5 in the main text. This may also be  relevant to studies of excited state  properties of black phosphorus.  Isosurfaces of wavefunctions at VBM and CBM at the X-point along with  a schematic diagram of projerction on p$_{x,y}$-orbitals of P showing symmetry of the wavefunction at VBM and CBM at X-point are shown in figure \ref{S5}.

\end{document}